\documentclass[conference]{IEEEtran}

\usepackage{url}
\usepackage[bookmarks=false]{hyperref}
\usepackage{cleveref}
\hypersetup{bookmarks=false}

\ifCLASSINFOpdf
  \usepackage[pdftex]{graphicx}
\else
\fi
\begin{document}
%
\title{Improving Big Data Visual Analytics with Interactive Virtual Reality}

\author{\IEEEauthorblockN{Andrew Moran}
\IEEEauthorblockA{Dept. of Electrical Engineering and Computer Science\\
Massachusetts Institute of Technology\\
Cambridge, Massachusetts\\
andrewmo@mit.edu}
\and
\IEEEauthorblockN{Vijay Gadepally, Matthew Hubbell, Jeremy Kepner}
\IEEEauthorblockA{Massachusetts Insitute of Technology\\
Lincoln Laboratory{\textsuperscript{1}} \\
Lexington, Massachusetts\\
\{vijayg, mhubbell, kepner\}@ll.mit.edu}}


%


\maketitle

\begin{abstract}

For decades, the growth and volume of digital data collection has made it challenging to digest large volumes of information and extract underlying structure.  Coined `Big Data', massive amounts of information has quite often been gathered inconsistently (e.g from many sources, of various forms, at different rates, etc.).  These factors impede the practices of not only processing data, but also analyzing and displaying it in an efficient manner to the user.  Many efforts have been completed in the data mining and visual analytics community to create effective ways to further improve analysis and achieve the knowledge desired for better understanding.  Our approach for improved big data visual analytics is two-fold, focusing on both visualization and interaction.  Given geo-tagged information, we are exploring the benefits of visualizing datasets in the original geospatial domain by utilizing a virtual reality platform.  After running proven analytics on the data, we intend to represent the information in a more realistic 3D setting, where analysts can achieve an enhanced situational awareness and rely on familiar perceptions to draw in-depth conclusions on the dataset.  In addition, developing a human-computer interface that responds to natural user actions and inputs creates a more intuitive environment.  Tasks can be performed to manipulate the dataset and allow users to dive deeper upon request, adhering to desired demands and intentions.  Due to the volume and popularity of social media, we developed a 3D tool visualizing Twitter on MIT's campus for analysis.  Utilizing emerging technologies of today to create a fully immersive tool that promotes visualization and interaction can help ease the process of understanding and representing big data.

\end{abstract}

\begin{keywords}
big data; Twitter; virtual reality; visual analytics; geospatial; 3D; visualization; data analysis; user interaction; situational awareness
\end{keywords}

\footnotetext[1]{This work is sponsored by the Assistant Secretary of Defense for Research \& Engineering under Air Force Contract \#FA8721-05-C-0002.  Opinions, interpretations, conclusions and recommendations are those of the author and are not necessarily endorsed by the United States Government.}

%
\IEEEpeerreviewmaketitle

\section{Introduction}

Information is continuing to accumulate and is being collected at an increasing rate. ``\textit{We are in The Age of Big Data}" \cite{lohr2012age}.  As of 2012, about 2.5 exabytes of data are created each day \cite{schroeck2012analytics}. Today, big data can be used to convey different concepts such as social media, marketing, financial services, advertising, etc. \cite{schroeck2012analytics}.  Much information can be used to characterize particular analytical models in practice; however, this massive intake of information can commonly be unstructured and overly complex.  In fact, the three main principles that govern Big Data include velocity, variety and volume \cite{mcafee2012big}.  These prime factors make it difficult to easily detect patterns and get an overall sense of the data's architecture.  Today's challenges are to develop meaningful tools for analysts and users to understand data in a more convincing way.

Visualization plays a key role in exploring and understanding large datasets.  Visual analytics is the science of analytical reasoning assisted by interactive user interfaces \cite{thomas2005illuminating}.  According to \cite{keim2002information}, there is much to gain when data is represented in a more visual way.  This capability will enable quicker time to insight and more direct interactions with information.  Big Data may contain certain anomalies and abstract features that are not so easily recognizable.  The goal of performing analytics is to uncover these underlying patterns and display it to the user effectively.  This exploration process of Big Data can be improved by integrating human intuition and perception.  Hence, the key concept of effective data visualization is to represent congested and complex data in a way that is more manageable for the user.

One strategy is to combine visual analytics with known geographical representations called geovisual analytics (GVA).  GVA describes the use of visuals with map-based interfaces to further support the understanding of information \cite{ho2008exploratory}.  The motive for GVA is to get a better sense of large datasets by having a contoured terrain in the background to help guide exploration and analysis.  As a result, users gain an additional sense of situational awareness by making comparisons and connections with their surroundings.  Geovisual analytics is also very helpful in determining patterns that may be better depicted when data can be geographically distributed.

When working in spatial and geographical domains, simulations and virtual reality can lead to better discovery.  Virtual reality (VR) has made many advances in the realm of game development by realistically reproducing first person perspectives \cite{wang2010new}.  Game engines such as Unity3D \cite{unity} are capable of constructing user experiences that connect computer graphics, interaction, creativity, etc. all together.  They have also been tested for applying techniques such as situational awareness \cite{hubbell2012large} and information visualization \cite{kot2005information}.  \cite{djorgovski2013mica} and \cite{donalek2014immersive} have shown how VR has extended from game applications into other areas of research.  These above works have demonstrated how immersion helps scientists more effectively investigate and perceive their area of study. Data visualization has shown to support analyses that are multi-dimensional and highly abstract.  According to the MICA experiment \cite{djorgovski2013mica}, utilizing virtual reality helps visualize and analyze large data in 3D space. \cite{donalek2014immersive} shows how VR can create a more collaborative and immersive platform for data visualization.  Utilizing VR technology as a data visualization tool is an emerging field of research with promising outlooks. 

Our approach is to develop a Unity3D application that takes advantage of geospatial visual analytics of Twitter data at MIT into a virtual reality setting.  Although the related social media work of MAPD \cite{mapd} and TwitterHitter \cite{white2010twitterhitter} are sufficient Twitter geo-analytical tools, they remain two-dimensional, revealing some limitations in clustering, aggregation, and perception.  By embedding catalogued tweets into a 3D geospatial environment, users can more directly perceive and interact with their data.

The remaining portions of this article is structured as follows. \Cref{implementation} describes the implementation of our application; from data extraction and pre-processing to game configuration and design.  \Cref{user_inter} discusses the user interaction our application provides; elaborating on the technologies used and the analytical tasks that can be performed by the user.  We provide a discussion of our results in \autoref{discussion}.  Finally, we conclude and mention areas of future work in \autoref{conclusion}.  

\section{Application Implementation}
\label{implementation}

At MIT Lincoln Laboratory, we have created a tool to help visualize geographical data for analysis.  We have juxtaposed thousands of geo-tagged Twitter tweets onto a 3D model of MIT's campus.  For data extraction, we utilized many tools available at the lab as described below.  The \textit{Unity3D}\textsuperscript{TM} game engine was used for visualization and will be described in further detail in \autoref{user_inter}.

\subsection{Data Extraction and Pre-Processing}

Developing an accurate geographical environment into a 3D simulation is important for user situational awareness and analysis.  Depending on the source, much pre-processing is involved to ensure optimal data is used for visualization and scene creation.  In the next subsections, we will describe two sources, LADAR and Twitter data.

\subsubsection{LADAR Data}\hspace{2pt}
\label{ladar}

As a sensing technology (developed at MIT Lincoln Laboratory), LADAR is utilized to generate 3D representations of global locations \cite{cho2006real}.  LADAR measures the distance of reflected light from a laser source to an illuminated target as an accurate metric for height mapping.  In 2005, a LADAR dataset was collected from an overhead aircraft over Cambridge, MA encompassing MIT's campus \cite{fetterman2014luminocity}.  With about 1m resolution, a dense height map was created where each planar point corresponds to the altitude at that location.  The final image resulted in a 1.0km x 0.56km region of Cambridge.  

To produce a 3D rendition of that particular region, the LADAR data is converted into a stereolithography STL file.  This is a common 3D file format that can be imported to various modeling programs for further customization and enhancement.  For noise reduction, 3D graphics and animation software such as \textit{Blender}\textsuperscript{TM} \cite{blender} and \textit{Maya}\textsuperscript{TM} \cite{maya} was used to smooth jagged vertices.  These were than exported to a FBX format so that it can be read into Unity.

Given this region of Cambridge, satellite imagery from Google Earth \cite{ge} provides additional context of the setting.  The longitudinal and latitudinal bounds of the area was (+42.350, -71.090) to (+42.357, -71.099).  Two square JPEG images, corresponding to about roughly one half km in world dimensions, were extracted from Google Earth to capture the entire scene.  These were then compressed to textures, each 2048 x 2048 pixels, to later be used in the game environment.

\subsubsection{Twitter Data}\hspace{2pt}
\label{twitterdata}

The Big Data source on which we wanted to perform further analysis is Twitter.  Twitter is a social media blogging site where users can post messages in the form of tweets \cite{kwak2010twitter}.  Analyzing tweets can help provide insight on social behaviors, controversial topics, user reputation, and popular locations.  If posted from a mobile device, tweets are bound with a geo-tagged location in addition to their username, text message, timestamp, etc.   These tweets are gathered from Twitter Decahose \cite{gnip}, which provides 10\% of random tweets, and can be narrowed down to user-defined criteria (e.g time and location).  

After ingesting raw data from Twitter Decahose, it is parsed into a tab separated value (TSV) format and stored on the high performance database (DB) Apache Accumulo \cite{apache}.  Using the same procedures exercised in \cite{weber2014using}, additional models can be used to further query the data.  Specifically, we utilized the Dynamic Distributed Dimensional Data Model (D4M) \cite{kepner2012dynamic}, a high performance schema that can be used with Accumulo. The D4M syntax allows for easy data filtering by latitude and longitude, as well as quickly inserting additional attributes to tweets that satisfy certain criteria 
(such as containing specified `buzz' words).  We extracted about 6,000 tweets over the course of five months from October 2013 - February 2014.  After filtering and processing the dataset, the remaining data is exported as a TSV file.

\begin{figure}[h]
\centering
\includegraphics[width=\linewidth]{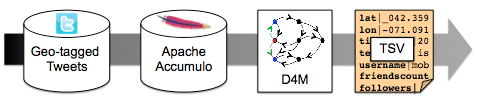}
\caption{Data Process Pipeline for Twitter Data.  Geo-tagged tweets are stored and pre-processed into a TSV format which can be parsed to render 3D objects in the scene.}
\label{pipe}
\end{figure}

\begin{figure*}[t]
\begin{center}
   \includegraphics[width=0.85\linewidth]{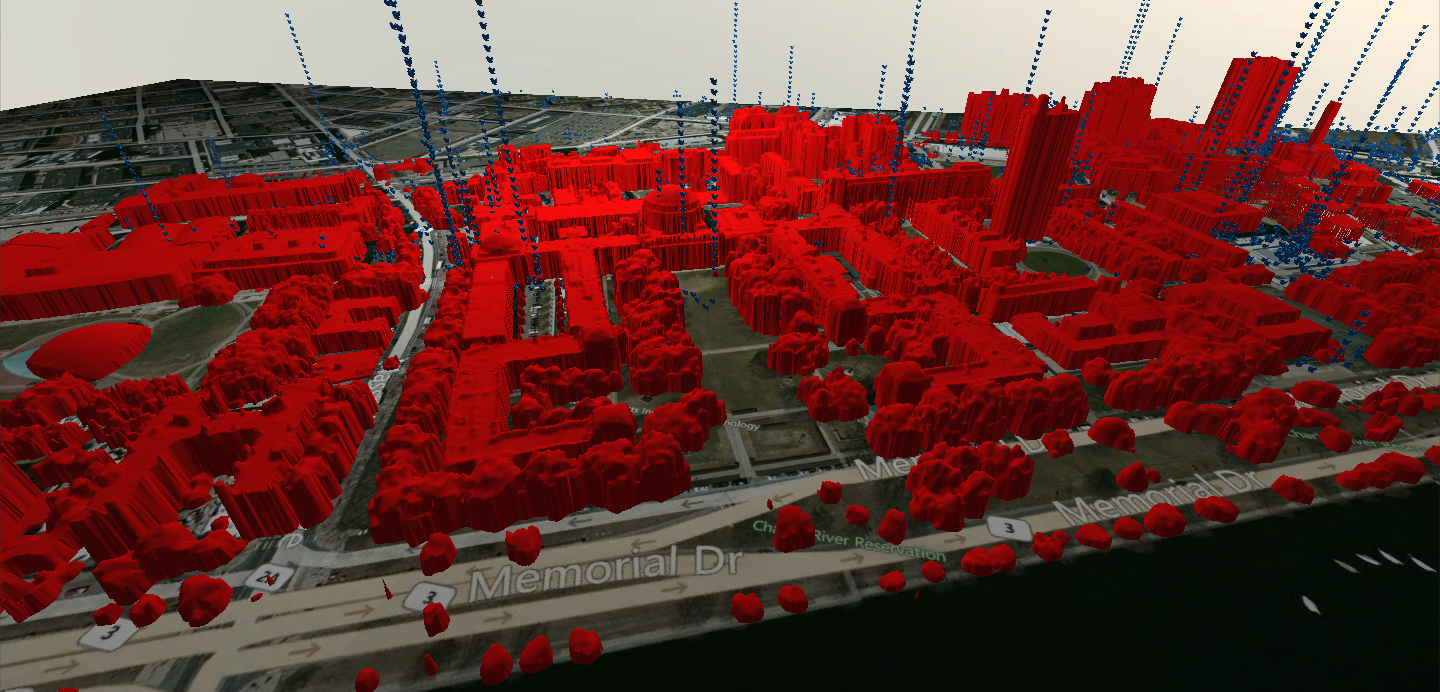}
\end{center}
   \caption{Rendition of MIT'€™s campus as an imported FBX model into the Unity3D engine as seen from game's free-form camera.  These models are then superimposed on Google Maps textures matching the same scale and latitude, longitude bounds as the original LADAR data.  Tweets are juxtaposed onto the scene based on provided mobile geo-tagged information.}
\label{fig:short}
\end{figure*}

\subsection{Configuration and Game Design}

After pre-processing the data, we now have formats that are easily imported into the Unity3D game engine.  Some manual configuration is necessary to ensure orientations and scales of landscapes appropriately create a realistic geography.

\subsubsection{Model and Scene Formation}\hspace{2pt}

Creating the static scene requires some manual configuration.  The textures provided from Google Earth were rendered on scaled 2D planes placed at the scene's origin.  Importing the LADAR FBX model into Unity produced several model subdivisions.  One constraint of Unity is that each imported model is limited to 65,000 vertices before partitioning itself into new models.  These models were arbitrarily sectioned and not necessarily positioned relative to the game's point of origin.  A global rotation and translation in the scene was performed on each section to properly connect models and ensure their positions matched correctly on the ground texture.  Similar to Google Earth and LADAR data collection, Unity's default unit is 1m. This made transitioning and manipulating elements in the scene very consistent.

Tweets additionally needed to be represented in the 3D world.  An open source delimited file reader was used to parse each tweet as an individual record, given a pre-defined header.  Of all the tweets, approximately 98\% were read in fully.  Ambiguous tweets that included unrecognized characters, invalid values, and/or missing fields were ignored.  Translating these records into the game environment required the use of publicly available models provided from Google Sketchup 3D Warehouse \cite{warehouse}. \textit{Maya}\textsuperscript{TM} \cite{maya} was used for further model enhancement and customization.  Attributes of each record coordinated which 3D model to use.  \Cref{query_pic} shows an example of how tweets are shown as blue birds by default whereas those containing the word ``danger" are represented as red skulls.  This corresponds to the result of a string matching analytic performed by D4M, as previously described in \autoref{twitterdata}.

Additional work was required to correctly map the geographical information provided by a tweet into the game world.  From \autoref{ladar}, the latitude and longitude boundaries of the LADAR and Google Earth images are well defined.  Therefore, translating real world latitude, longitude locations to game coordinates required a simple geometric transformation onto the scene's game ground layer.

\subsubsection{Game Elements and User Interface}\hspace{2pt}

After the static scene has been configured, additional elements are implemented in the environment to enhance immersive gameplay and promote visual analytics.

Initially, the user is instantiated as a first person controller.  With a free-form camera, the player's perspective can dynamically change in the x,y,z directions and is free to navigate within the bounds of the scene.  Colliders on buildings, tweets, and other 3D models prevent the player from reaching areas with obstructed views within objects. 

To confirm player direction and orientation, a 3D cursor/crosshair is shown on a transparent texture in front of the player's camera.  This is used to also help pinpoint where on the 3D scene the player is currently looking and facing.  As the player is constantly moving, the cursor remains in the center of the screen.  If the user chooses to pause player movement, the cursor is no longer fixed and is free to interact with game elements within the camera's current field of view.  

\begin{figure}[h]
\centering
\includegraphics[width=.95\linewidth]{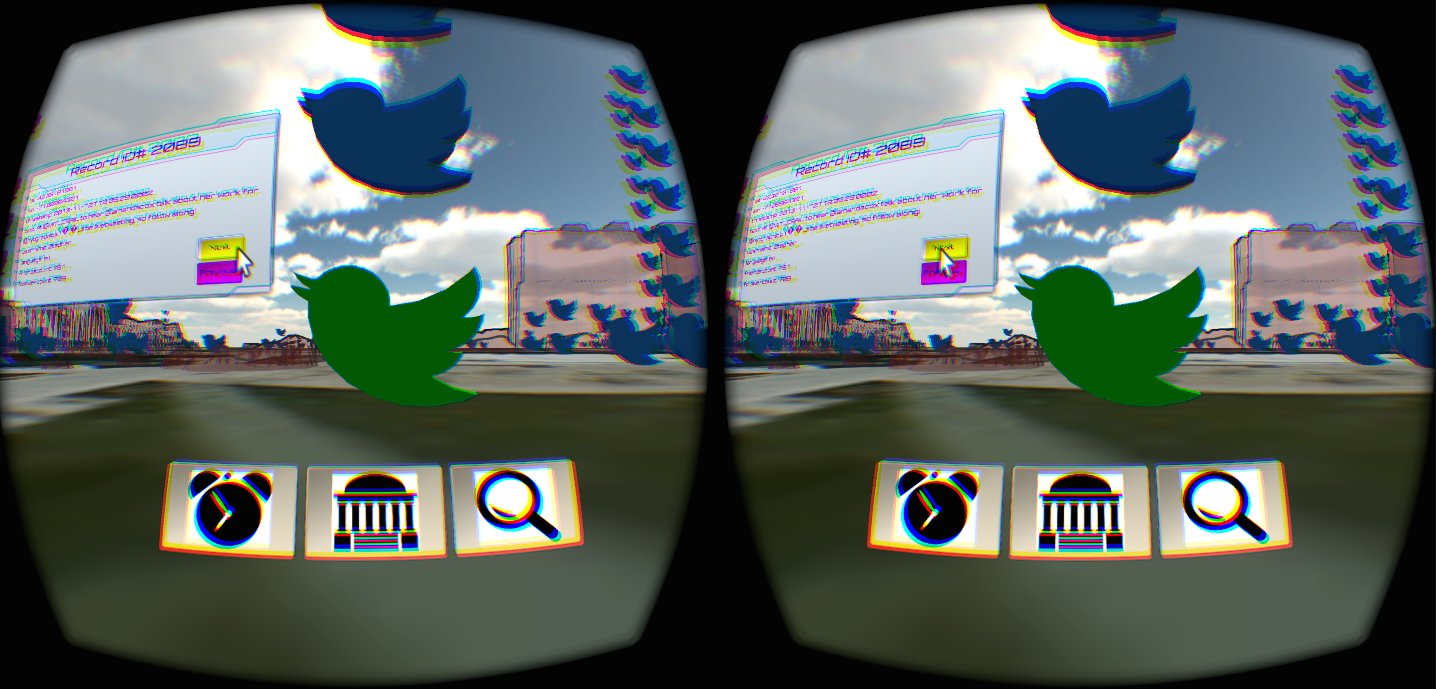}
\caption{View of selected tweet and HUD as seen from the stereoscopic view of the Oculus Rift, a VR device described in \autoref{techno}.  Utilization of 3D space allows freedom of GUI placement; whether at a fixed distance in front of the player or on 3D objects }
\label{stereo}
\end{figure}

As shown in \autoref{stereo}, additional GUI elements on the Heads Up Display (HUD) are displayed to help guide the player into further investigation on the Twitter dataset.  Current options included filtering time ranges, changing object opacities, and performing searches on the Twitter dataset.  These analytical tasks describing filtering and queries are described in \autoref{an_tools}.   

\section{User Interaction}
\label{user_inter}

With a game-like simulation generated, the player is ``more involved" in the scene.  To promote analysis, interaction is necessary to promote cognitive understanding and quickness to insight.

\subsection{Combining Technologies}
\label{techno}

This project embedded information from large datasets into the \textit{Unity3D}\textsuperscript{TM} game engine \cite{unity}.  Unity3D is a fully capable physics engine that is readily available to developers and highly reputable in performance.  Its flexibility in multi-platform support and scripting makes it a valid candidate as a modeling and 3D visualization tool.  \cite{djorgovski2013mica} and \cite{donalek2014immersive} show examples of how Unity3D is extending its visualization as an emerging development tool for virtual reality.

Unity3D also integrates software development kits (SDKs) for various hardware specialized in collecting data from realtime user input.  Tools such as the \textit{Oculus Rift}\textsuperscript{TM} \cite{oculus} and \textit{Leap Motion}\textsuperscript{TM} \cite{leap} consist of sensors, cameras, positional tracking and enhanced displays to record player inputs and directly relay that information in the 3D setting.  Combining these commercial yet portable technologies will help maximize the interaction and immersion we want to perceive in the 3D geographical data representation. 

\begin{figure}[h]
\centering
\includegraphics[width=0.5\linewidth]{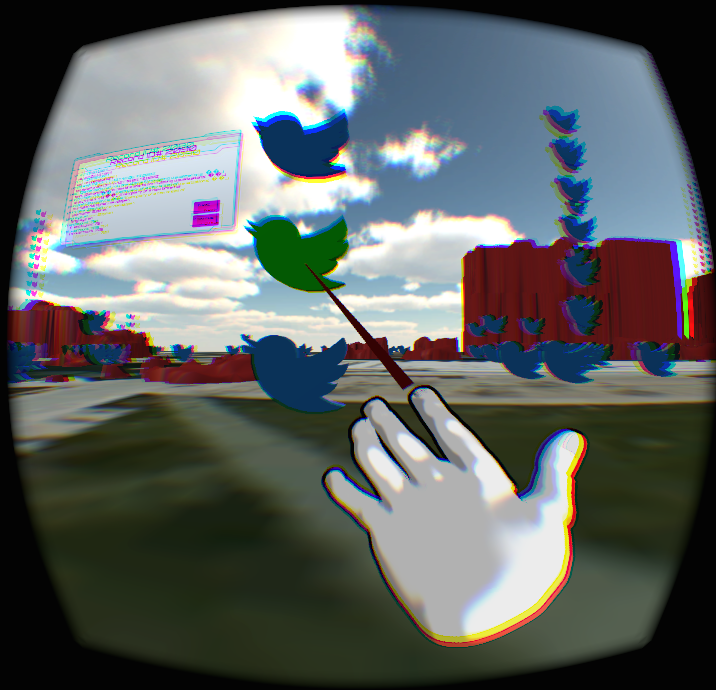}
\caption{Leap Motion hand controller allows the player's hands to be rendered in the simulated scene.  Gestures and other inputs registered by the device can launch events and other commands intended by the user during analysis and gameplay.}
\label{gesture}
\end{figure}

\subsection{Analytical Tasks}
\label{an_tools}

Interaction techniques fuse together user input with output to provide a better way for a user to perform a task \cite{tasks}.  Common tasks that allow users to gain a better understanding of data include scalable zooms, dynamic filtering, and annotation.  Below, we describe some tasks that can be performed fluidly by the user and how an enhanced situational awareness is achieved in our application of the MIT Twitter dataset. 

\subsubsection{Navigation/Exploration}\hspace{2pt}

Creating a life size simulated setting enables the player to naturally move about the scene.  Virtual reality fully immerses the player and enables a constant stimuli for exploration and discovery.  Using MIT's campus allows players to recognize familiar landmarks and discover new regions of interest (ROI).  Utilizing a free-form camera permits different perspectives that would not have been so credible in the real world.  Adjustable zooming is possible by having the camera move closer or farther from a relative position in the scene.  The user's freedom to move about the 3D scene is key to revealing the overall framework and features of the dataset which would not have been so noticeable in a traditional display.

\subsubsection{Identification/Selection}\hspace{2pt}

\begin{figure}[h]
\centering
\includegraphics[width=.95\linewidth]{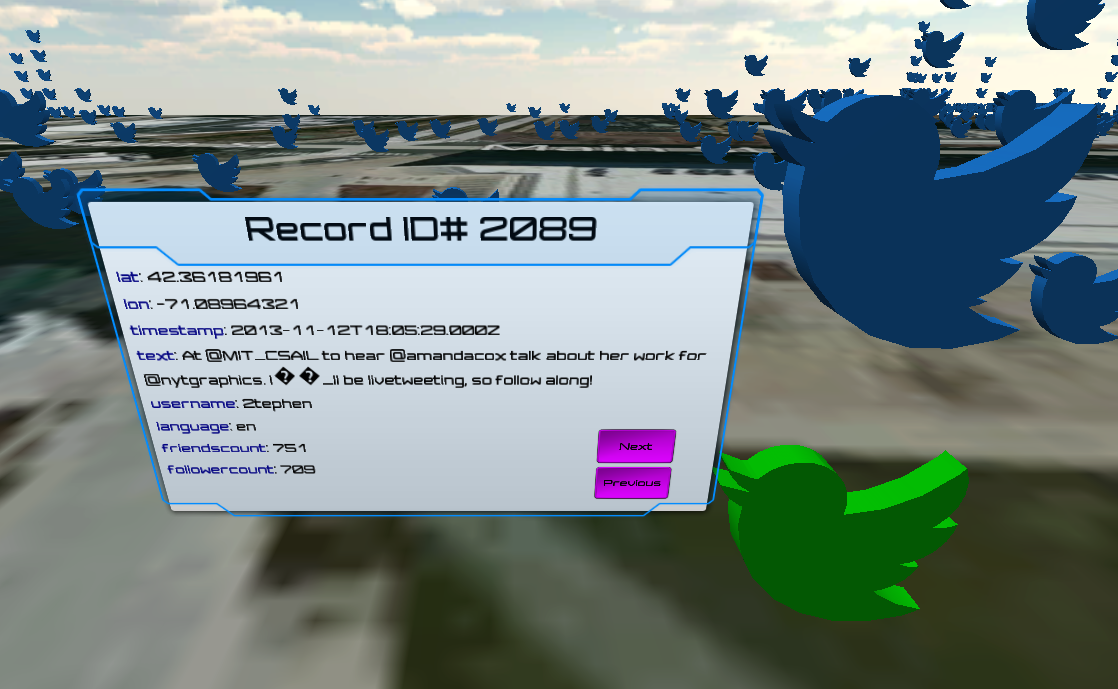}
\caption{Upon selection, the 3D representation of a tweet changes color and launches a speech bubble revealing all characteristics.}
\label{speech}
\end{figure}

Tweets are represented as 3D objects in the environment.  The status of a tweet can be represented visually by the model observed by the player.  Characteristics of tweet models such as type, size, color and motion allow the player to instantly know the nature of the tweet.  These visual queues now give the player an enhanced situational awareness.  Users have the option to perform actions on their setting to further dive deeper into the dataset.  As shown in \autoref{speech}, a player can select a tweet revealing a 3D display showing all the original data as it was read in such as username, follower count, timestamp, text, etc.      

\subsubsection{Filtering/Dynamic Queries}\hspace{2pt}

\begin{figure}[h]
\centering
\includegraphics[width=.95\linewidth]{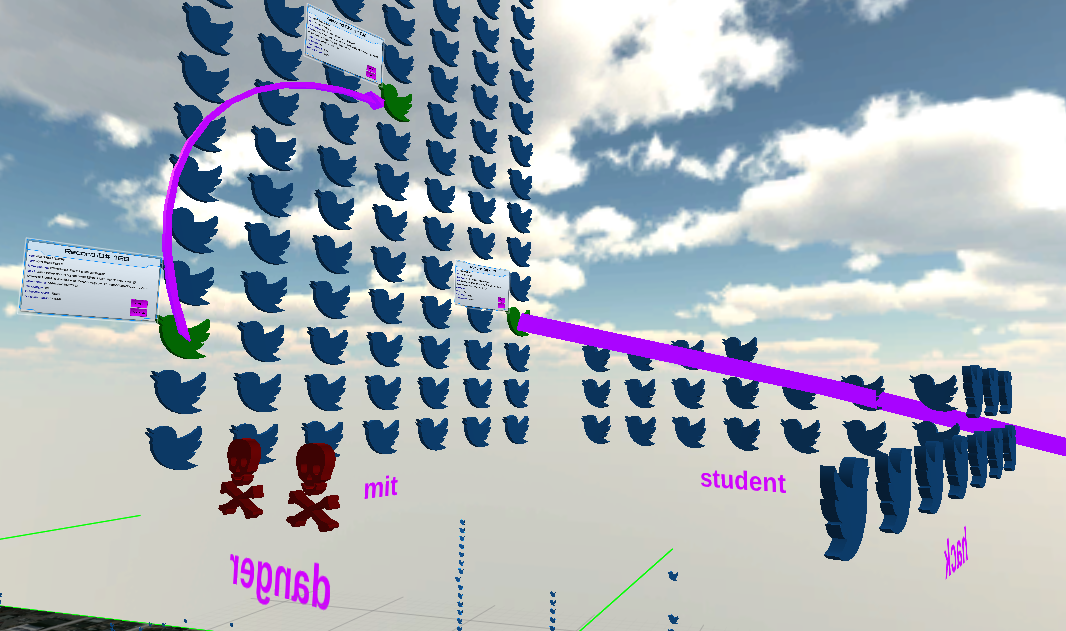}
\caption{ Queries can be performed on the dataset to create a floating virtual room where walls are populated by tweets that match user defined criteria.  }
\label{query_pic}
\end{figure}

As shown in \autoref{stereo}, menu options on the GUI allows for further analysis on the data.  Being able to apply filters and dynamic queries can help analysts focus on specific features, reveal underlying structure, and formulate hypotheses.  In this project, there are a few ways in which we can filter the Twitter data.  Analysts can select a time interval in which the tweets were timestamped to narrow down the dataset within a preferred range.  Another option is to change the physical landscape by adjusting the opacity of buildings rendered in the scene.  By default, buildings are fully colored.  However, there are options to change shaders applied to the 3D model such that it is wire-framed or completely transparent.  This allows the option to compare tweets in separation or in conjunction with their landscape.  Additionally, tweets can be searched by keywords that produce groups in three-dimensional space.  By use of a virtual keyboard, users type and define a criteria to do a string match on the tweets.  If a match exists, the tweet moves from its original location to a new one where a virtual wall is formed as shown in \autoref{query_pic}.  This allows the analysts to see connections and relationships between various Twitter topics, locations, users, etc.

\subsubsection{Clustering/Pattern Recognition}\hspace{2pt}

Overlaying data on top of the geographical landscape in which it was produced can make it easier to detect patterns.  For example, some tweets in this dataset share common characteristics such as location, topic, etc.  In the default physical view, if a user posts a tweet at the same location of another one, the new tweet is physically placed on top of the previous tweet.  As a result, vertical stacks can be created in the environment where the ordering is shown chronologically from bottom to top in which the Twitter posts have been timestamped.  This clustering can help define the nature of the geography or the social behavior of users.  For example, clusters can be seen around popular public places such as dining halls and dormitories on MIT's campus and less on the academic side of campus.  Another noticeable pattern is that some individual users post in bursts in which they make multiple tweets from the same location.

\subsubsection{Detail-On-Demand}\hspace{2pt}

With a tweet of interest, additional actions can be performed to reveal new information specific to that particular tweet.  Hovering over and selecting the tweet with a virtual cursor opens a display in 3D space. As shown in \autoref{speech}, we can see all the attributes that are associated to that tweet when it was initially read into the database.  Additionally, there are other actions that can be performed to help track user behavior. One option is to show a user's preceding or succeeding tweet if there exists one in the dataset.  This renders a directed 3D waypoint arrow in the scene revealing the user's next location at which they made a tweet, relative to their previous post.  This helps show routes of users and known behaviors given geographical information.    

\begin{figure}[h]
\centering
\includegraphics[width=.95\linewidth]{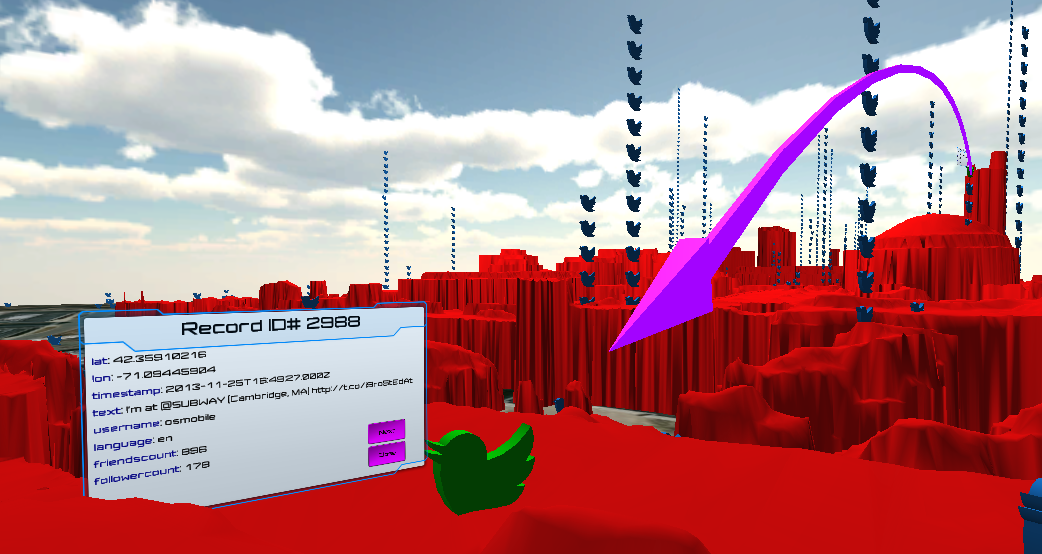}
\caption{Waypoint arrows rendered in the virtual world lets the player track social behavior of Twitter users in the order in which the tweet was delivered.  }
\label{way}
\end{figure}

\section{Discussion}
\label{discussion}

One of the main challenges with Big Data today is coming up with a proper representation to the user for effective analysis.  As data scales into higher dimensions, it can become overly complex.  Visualization is key in the aid of pattern recognition and data analysis.  At Lincoln, we experimented with using novel methods and emerging technologies of today to enhance visualization and user interaction for data analysis.  Virtual reality creates an immersive environment for the user such that as data is overlaid within a geographical domain, an enhanced situational awareness and cognition was achieved.  

These advances in virtual reality continues to grow as computation and processing becomes faster on both the hardware and software fronts.  As a result, these devices are becoming more powerful, affordable, and readily available to the research and development community.  This increases the capability of integrating visual data exploration and interaction within virtual reality.

\begin{figure}[h]
\centering
\includegraphics[width=.7\linewidth, trim=0 0 480 0, clip]{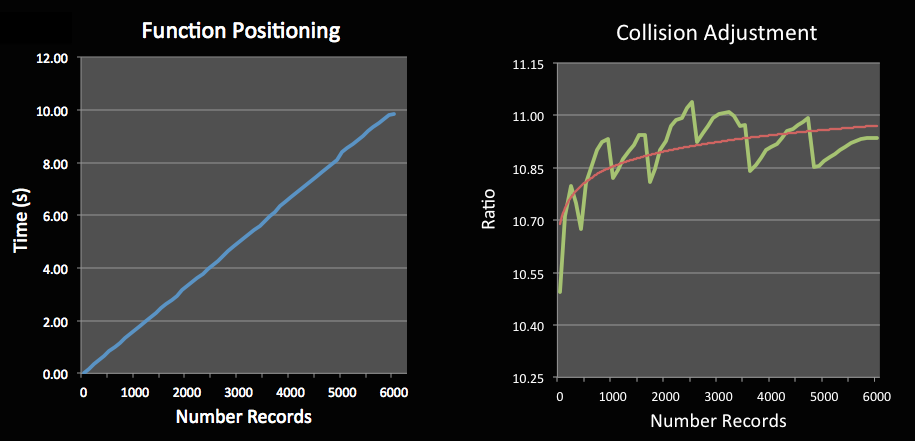}
\includegraphics[width=.7\linewidth, trim=470 0 10 0, clip]{pics/graphs.png}
\caption{Top: Linear progression comparing time to complete function call for positioning tweets with number of tweets present. Bottom: Logistic progression comparing ratio of number of collisions with number of tweets. }
\label{graphs}
\end{figure}

Performance and a high frame rate is important when working in simulations that show many data points.  Ingesting the data on Accumulo with D4M analytics is proven to be fast.  D4M achieved 100,000,000 inserts per second as it's peak performance \cite{kepner2012dynamic}.  Most of the computation comes from parsing the pre-processed Twitter data and constructing the 3D scene layout.  Collision detection amongst tweets on instantiation requires a considerable amount of computation.  \Cref{graphs} shows how performance of positioning tweets is effected once the game starts with the new data.

For demo and portability interests, this work has been completed on a Macbook laptop.  Although producing promising results, there were some foreseeable limitations.  As more objects populate the scene, more system checks are completed frame by frame.  It is recommended to have a faster processor to achieve better performance and reduce jerky movement as the camera pans a scene (e.g. scene judder).  Oculus suggests a frame rate of at least 60-75 fps for a comfortable user experience.  With more vertices rendered in the scene, more draw calls are sent to the GPU.  In addition, Oculus Rift rendering and Leap Motion gesture recognition requires a lot of processing.  Upgrading from a traditional laptop to a more powerful machine can produce a more ideal game experience.

\section{Conclusion and Future Work}
\label{conclusion}

Although much progress has been made, further improvements could enhance both application performance and user interactive gameplay.  Rendering 3D models scales linearly with performance.  Activating and deactivating colliders when needed can help reduce the computation load.  Additional shaders could be applied to the 3D buildings of Cambridge to provide a better rendition and give the player more options of how the tweets are overlaid in the scene.  Occlusion layers for overlapping tweets and blocked buildings could be applied to prevent unnecessary rendering.

Although we have a few useful analytics now, we intend to add more features that allow for further engagement by the player.  Originally, this work was done in Unity 4.6 and Oculus Rift DK1.  Implementing Unity's UI system allows for 3D text and more engagement with Leap Motion.  Continuing to exercise 3D interactions from hand inputs rather than gamepad controllers could help immerse the player and manipulate the data more effectively.  Currently, we are transitioning to the Oculus Rift DK2 to utilize the enhanced display and accurate positional tracking.  Some potential future features we plan to implement in the user-interface include multi-selection and annotation.  We also plan to continue researching other ways to enable user interaction and improve usability.  

This project reveals the added potential of how utilizing the VR platform can bring a more effective visual experience.  We have effectively visualized Twitter on a 3D model of MIT's campus to improve Big Data visual analytics.  This research has shown how (1) virtual reality can also be used as a data visualization platform, (2) a more immersive environment enables user interaction, (3) patterns and visual analytics are more efficient when working in a geospatial domain.


%



%



{\small
\bibliographystyle{ieeetr}
\bibliography{bare_conf}
}

\end{document}